 \documentclass[final,5p,times,twocolumn,authoryear]{elsarticle}

\usepackage{amssymb}
\usepackage{lipsum}
\usepackage{tabularx, makecell, booktabs}
\usepackage{rotating}
\usepackage{graphicx}%
\usepackage{amsmath}%
\usepackage{amsfonts}
\usepackage{xcolor}
\usepackage{amsthm}%
\newcommand{\farcs}{\hbox{$.\!\!^{\prime\prime}$}}

\newcommand{\kms}{km\,s$^{-1}$}

\journal{New Astronomy}

\begin{document}

\begin{frontmatter}



\title{Insights on Gas Distribution and Dynamics in Massive Proto-cluster G358.46$-$0.39: Possible Multiplicity in G358.46$-$0.39 MM1a}


\author[a]{Chukwuebuka J. Ugwu}
\author[a]{James O. Chibueze}
\author[a]{Willice Obonyo}
\author[b]{Mavis Seidu}
\affiliation[a]{organization={UNISA Centre for Astrophysics and Space Sciences (UCASS), College of Science, Engineering and Technology, University of South Africa},%Department and Organization
            addressline={Cnr Christian de Wet Rd and Pioneer Avenue}, 
            city={Florida 1709},
            postcode={P.O. Box 392, 0003 UNISA}, 
            country={South Africa}}
            
\affiliation[b]{organization={Centre for Space Research},%Department and Organization
            addressline={North-West University}, 
            city={Potchefstroom},
            postcode={2520}, 
            country={South Africa}}

\begin{abstract}
This work explored the spatial distribution of C$^{17}$O, SiO, HC$_{3}$N and SO$_{2}$ molecules, as well as the energetics of outflows in G358.46$-$0.39 proto-cluster using ALMA band 7 archival data, with the aim of providing an improved understanding of its protostellar nature, gas kinematics and dynamics. G358.46$-$0.39 is previously known to consist of 4 dust continuum cores (MM1a, MM1b, MM1c and MM2). The integrated intensity map of C$^{17}$O reveals filamentary and dumbbell-shaped structures that are probably compressed gases from the expansion of the HII region MM2. The SiO emission reveals spatially overlapped blue and red outflow lobes, likely driven by an unresolved young stellar object (YSO) in MM1a. The spatial distribution of HC$_{3}$N and SO$_{2}$ molecules in MM1a shows a compact morphology, with no detectable HC$_{3}$N and SO$_{2}$ emissions in the other cores. The SO$_{2}$ emission reveals a clear velocity gradient in MM1a, as well as large velocity dispersion ($\sim$ 3\,\kms) within the inner core of MM1a, which are consistent with rotating structures. We estimated the mass, momentum and energy outflow rate, as well as other outflow parameters. The SiO outflow exhibits a different morphology compared to the $^{12}$CO outflow morphology previously observed in MM1a. The SiO and $^{12}$CO outflows are probably associated with disks of separate cores with one face-on and the other edge-on, pointing to multiplicity of YSOs in MM1a. The properties of MM1a indicate that it is a massive protostar that is actively accreting and undergoing star formation.
\end{abstract}




\end{frontmatter}

%\tableofcontents

%\linenumbers


\section{Introduction}
\label{introduction}

In spite of the key role high-mass stars ($\geq 8\,M\odot$) play in evolution of galaxy and shaping of the interstellar medium (ISM) through massive outflows, dissociative ultraviolet (UV) radiation, stellar winds and expanding HII region, their early formation stage remains poorly understood because of their complex environment \citep[e.g.][]{2012ARAandA..50..531K}. High-mass stars form in very obscured environment (deeply embedded), making them very difficult to be observed in their early phases at optical and infrared wavelengths. These objects have short evolutionary lifetimes or their formation phases occur quickly \citep[$\sim$ 10$^5$\,yr;][]{2021AandA...652A..71S}. They form in distant clusters and associations, therefore it is hard to isolate single high-mass star observationally \citep[e.g.][]{2007ARAandA..45..565M,2007ARAandA..45..481Z,2018ARAandA..56...41M,2025ARAandA..63....1B}. These are what have placed high-mass star formation in the front burner of astrophysics research.

Outflows and rotating structures give information about the kinematics of circumstellar gas in YSOs and studying these features (outflows and rotating structures) will provide improved understanding of the early formative stage of these objects \citep[e.g.][]{1983ApJ...265..824B,2003MNRAS.339.1011G,2007prpl.conf..197C,2007ApJ...669..464S,2007AandA...470..269Z,2009ApJ...707....1B,2010MNRAS.404..661V,2012ApJ...748..146C,2018ApJ...860..119G,2018AandA...620A.182K,2019PASJ...71...44H,2020SSRv..216...43Z}. Outflows are known to be associated with violent ejections of high velocity gas that arise in two directions from the central protostar, with enormous energy comparable to those seen in accretion events \citep{1996ARAandA..34..111B,2009ASSP...13..381B,2009AandA...499..811L}. Bipolar outflows are primarily driven by YSOs that are ingrained in their natal environment. The processes of discharging gas as outflows and jets are gradual but effective means of eliminating surplus angular momentum from the accretion disks, thereby enabling accretion to continue to take place \citep{1996ARAandA..34..111B,2009ASSP...13..381B}. Moreso, outflows and disks reveal information about the kinematics of circumstellar gas in YSOs. Some of the molecular line emissions such as CO and CH$_{3}$OH observed at submillimeter and millimeter wavelengths are good tools for exploring the physical processes (such as outflows, infall/collapse motions, jets and accretion burst) of ongoing star forming regions \citep[e.g.][]{2018ApJ...860..119G, 2019AandA...632A..57C,2019AandA...623A..77S,2020ApJ...896..127K,2025MNRAS.539..145C}. G358.46$-$0.39 has an accretion disk and associated outflow and thus provides important information on accretion ejection dynamics in star formation. It is an excellent laboratory for examining the nature of the accretion process (monolithic or competitive), which is a fundamental but poorly understood process in the formation of high-mass stars.

G358.46$-$0.39 is a periodic 6.7 GHz Class II CH$_{3}$OH maser source \citep{2015MNRAS.446.2730M} and a massive proto-cluster \citep{2017ApJ...836...59C,2017AandA...600L..10C}, that is associated with outflows \citep{2013ApJS..206...22C} and rotating envelope or disk \citep{2023MNRAS.520.4747U}. It is located at a kinematic distance of 3.9\,kpc and has multiple dust continuum cores (MM1a, MM1b, MM1c and MM2), with different millimeter spectra \citep{2023MNRAS.520.4747U}. It has a bolometric luminosity of $\sim$ 1.15 $\times$ 10$^{4}$$L_{\odot}$ and an approximate overall mass of 695$M_{\odot}$ \citep{2023MNRAS.520.4747U}. A deep look into the gas distribution and outflows in the source has not been properly explored. This study provides a more detailed investigation into the spatial distribution of specific molecules (such as C$^{17}$O, SiO, HC$_{3}$N and SO$_{2}$) and outflows in the source. These molecular line tracers are sensitive to different conditions, such as dense (C$^{17}$O, HC$_{3}$N and SO$_{2}$), shocked (SiO) and hot (HC$_{3}$N and SO$_{2}$) gas. As such, the molecules are ideal for a comprehensive probe of the different physical and chemical environments within G358.46$-$0.39. The study of gas motions (spatial distribution and kinematics) and the dynamics of gas in the vicinity of massive young stellar objects (MYSOs) would be very significant in tackling the mystery of early massive protostar formation. This work will not only provide significant insight into the gas distribution and outflows in the source, but will also contribute to the growing statistics of well-studied MYSOs. 

%It is located at a near kinematic distance of 3.9\,kpc \citep{2023MNRAS.520.4747U}. \citet{2023MNRAS.520.4747U} disentangled the source into 4 dust continuum cores (MM1a, MM1b, MM1c and MM2), revealing different millimeter spectra. The authors showed that MM1a exhibits copious hot core line emissions and is associated with highly collimated bipolar outflow traced by $^{12}$CO emission, as well as rotating structures (envelope or disk) traced by C$^{17}$O and CH$_{3}$OH emissions.

This work explores the spatial distribution of C$^{17}$O, SiO, HC$_{3}$N and SO$_{2}$ molecules, as well as the energetics of $^{12}$CO and SiO outflows in G358.46$-$0.39 proto-cluster using Atacama Large Millimeter/Submillimeter Array (ALMA) band 7 archival data in order to enhance our understanding of its protostellar nature, gas kinematics and dynamics. We describe the archival ALMA data used in this study in Section\,\ref{sec:observation} and present the results of the gas distribution in Section\,\ref{sec:result}. The discussions on the derived outflow properties compared to other works and the effect of disk orientation on outflow morphology are presented in Section\,\ref{sec:discussion}, while the conclusions from our findings are presented in Section\,\ref{sec:conclusion}.

\section{Observation}
\label{sec:observation}
%\lipsum[1]

Archival ALMA band 7 data of G358.46$-$0.39 with project code 2013.1.00960.S (PI: Csengeri T.) was employed in this work. The data were downloaded from the ALMA Science Archive\footnote{http://almascience.eso.org/aq/} and analyzed with the aid of the Common Astronomy Software Applications ($\mathrm{CASA}$) package \citep{2007ASPC..376..127M}. Our specific molecular lines of interest in the observations are C$^{17}$O (J = 3 $-$ 2) at 337.061\,GHz, SiO (J = 8 $-$ 7) at 347.330\,GHz, HC$_{3}$N (J = 37 $-$ 36) at 336.520\,GHz, SO$_{2}$ (19$_{1, 19}$ $-$ 18$_{0, 18}$) at 346.652\,GHz and $^{12}$CO (J = 3 $-$ 2) at 345.795\,GHz. The basics of the observed lines are listed in Table\,\ref{tab:properties}. The angular resolution of the observations is 0\farcs9 $\times$ 0\farcs6. The frequency and velocity resolutions are $\sim$ 0.977\,MHz and 0.879\,kms$^{-1}$, respectively. Full details of the observations, calibration and imaging are presented in \citet{2017AandA...600L..10C} and \citet{2023MNRAS.520.4747U}. 

 The C$^{17}$O, SiO, HC$_{3}$N, SO$_{2}$ and $^{12}$CO cubes were created using $\mathrm{CASA}$ task $\mathrm{imsubimage}$. Continuum subtraction was carried out on the C$^{17}$O, SiO, HC$_{3}$N, SO$_{2}$ and $^{12}$CO cubes, using CASA task $\mathrm{imcontsub}$ to eliminate continuum emission contribution on the molecular lines. Doppler correction was applied by re-framing the velocities of the C$^{17}$O, SiO, HC$_{3}$N, SO$_{2}$ and $^{12}$CO to the local standard of rest kinematics (LSRK), with respect to the rest/central frequency of the molecular lines, using CASA task $\mathrm{imreframe}$. The integrated intensity maps (moment 0) of C$^{17}$O, SiO, HC$_{3}$N and SO$_{2}$ emissions and the velocity field maps (moment 1) of HC$_{3}$N and SO$_{2}$ emissions, as well as the velocity dispersion map (moment 2) of SO$_{2}$ emission were formed using CASA task $\mathrm{immoments}$. The spectra of the molecular lines were extracted at the position of the continuum source, using spectral profile tool in $\mathrm{CASA}$. The moment maps and spectral plots were displayed using Astronomical Plotting Library in Python ($\mathrm{APLpy}$) and $\mathrm{Matplotlib}$, respectively.

\begin{figure*}
\centering
%\begin{tabular}{lc}
	\includegraphics[width=0.81\textwidth]{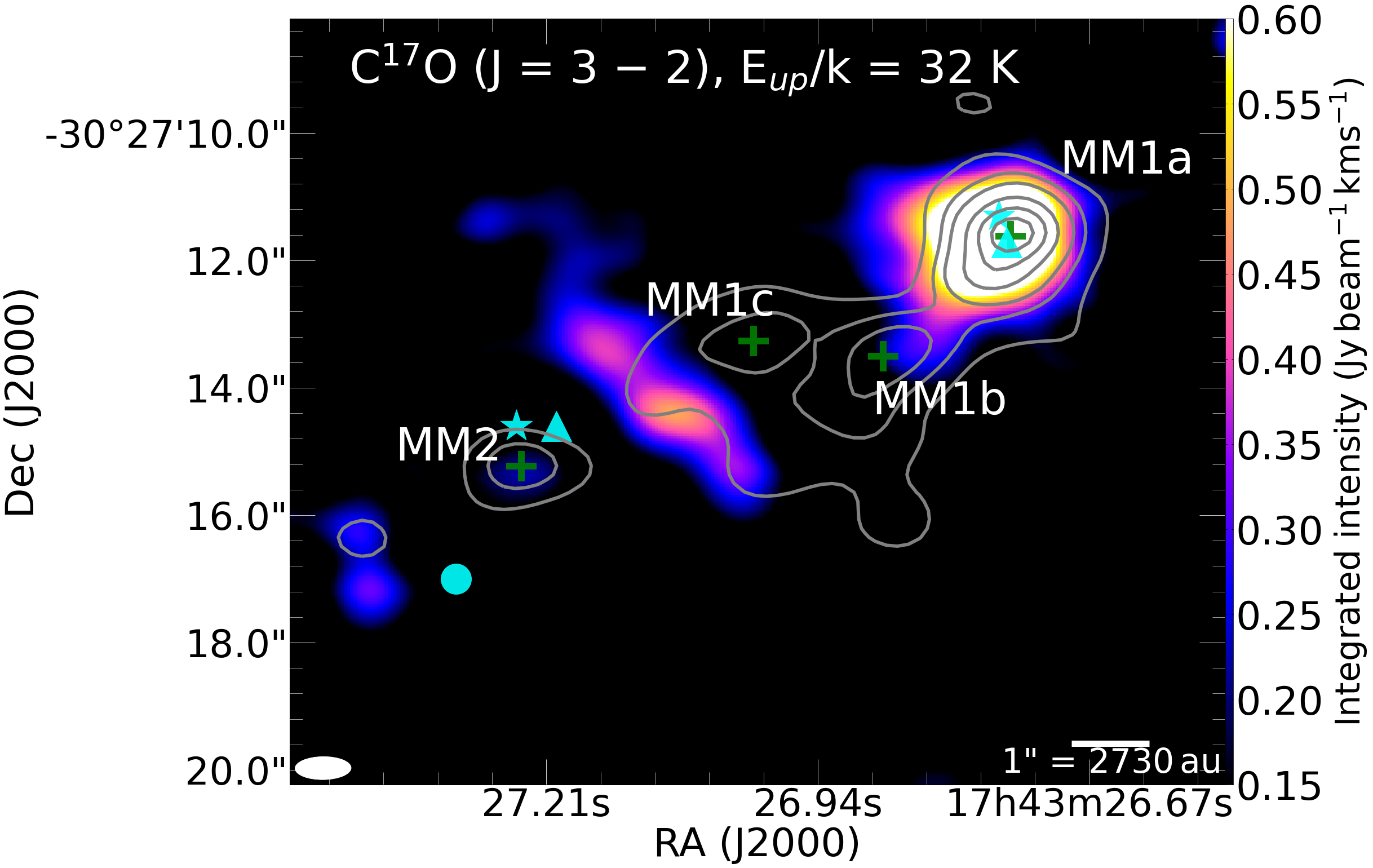} 
    \includegraphics[width=0.8\textwidth]{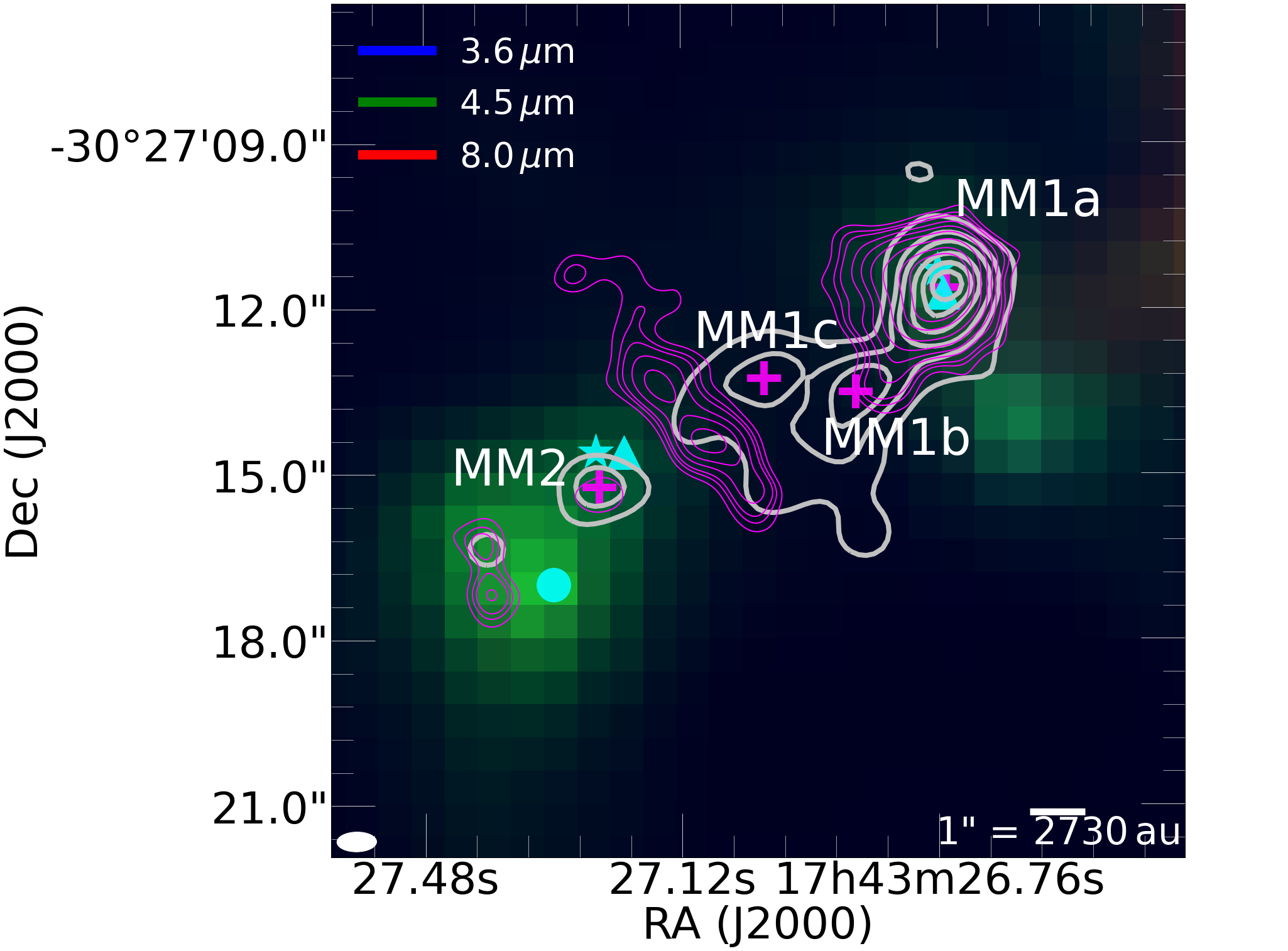}
 %   \end{tabular}
    \caption{$Top$: Integrated intensity map of C$^{17}$O emission (Background). $Bottom$: Integrated intensity map of C$^{17}$O (magenta contours with levels = 0.18, 0.23, 0.26, 0.32, 0.37, 0.44, 0.56, 0.71, 0.83\,Jy\,beam$^{-1}$\,kms$^{-1}$) and ALMA dust continuum image \citep[Gray contours with levels = 15, 30, 45, 75, 105, 135\,mJy\,beam$^{-1}$][]{2023MNRAS.520.4747U} superimposed on the Spitzer three-colour composite image \citep{2003PASP..115..953B}. Green and magenta crosses show the peak position of the dust continuum cores. Cyan stars, triangles and circle represent the peak position of the 6.7\,GHz CH$_{3}$OH \citep{2010MNRAS.404.1029C}, 22\,GHz H$_2$O \citep{2014MNRAS.442.2240W} and 95\,GHz CH$_{3}$OH \citep{2017ApJS..231...20Y} masers, respectively. White ellipse at the left bottom corner represents the ALMA synthesized beam.}
    \label{fig:c17o_m0}
\end{figure*}

\begin{figure*}
\centering
 \includegraphics[width=0.9\textwidth]{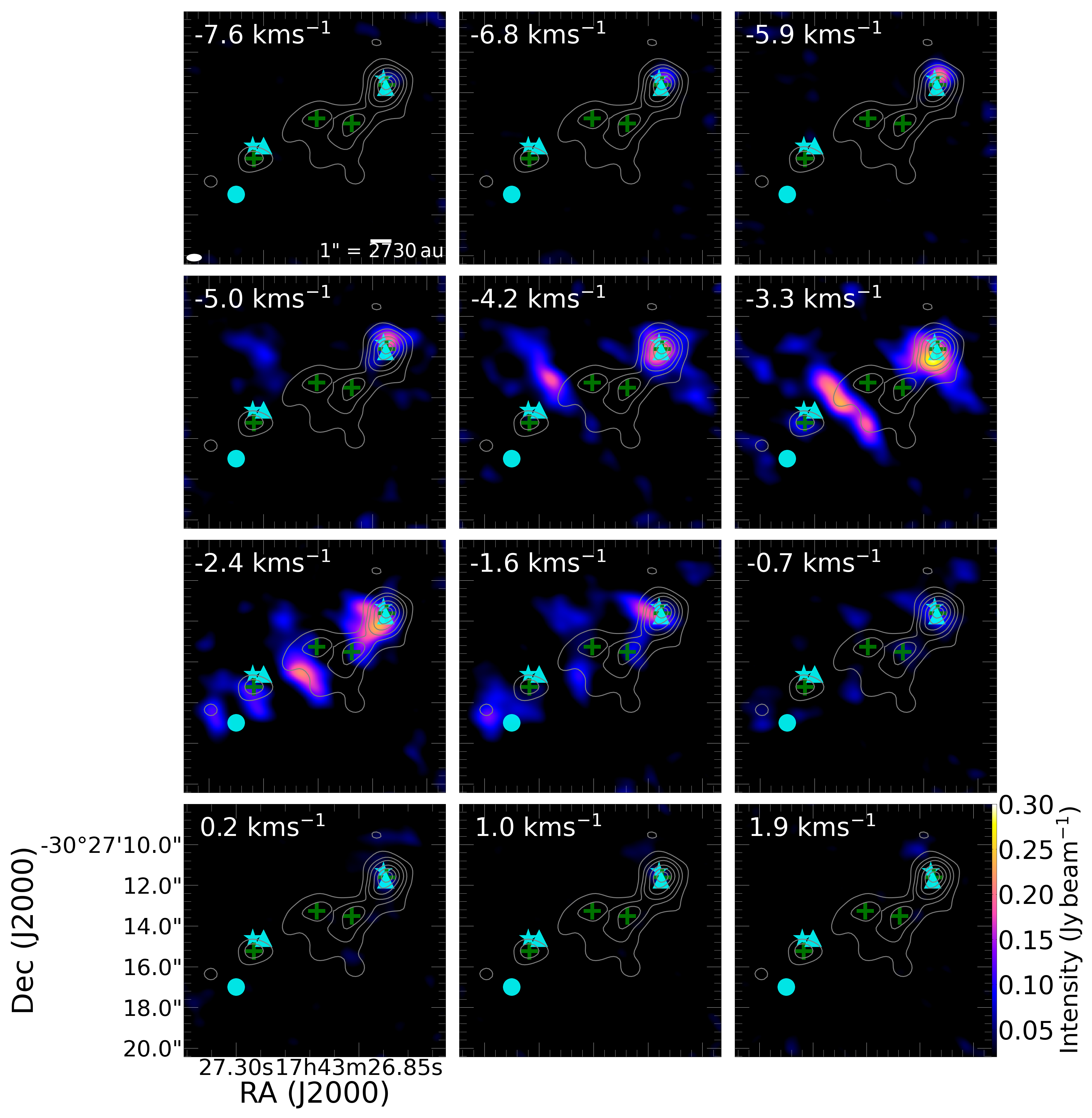}
   \caption{C$^{17}$O emission channel maps. Gray contours, green crosses, cyan stars, triangles and circle are the same as defined in Fig.\,\ref{fig:c17o_m0}.}
    \label{fig:C17O_chanmaps}
\end{figure*}

\begin{figure*}
\centering
 \begin{tabular}{cl}
	\includegraphics[width=0.5\textwidth]{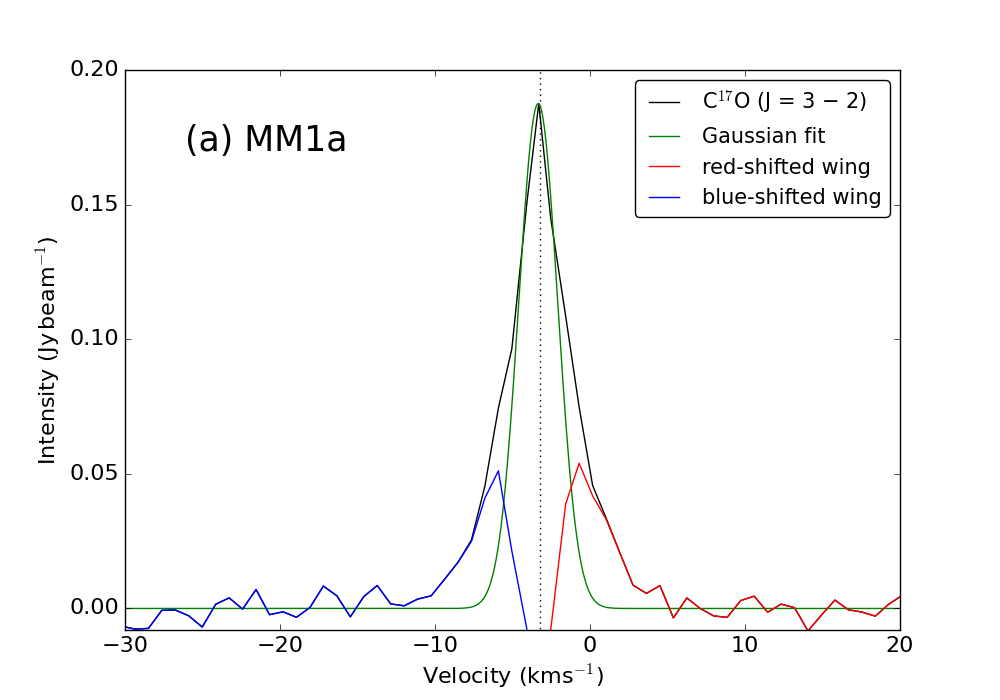} &
 \includegraphics[width=0.5\textwidth]{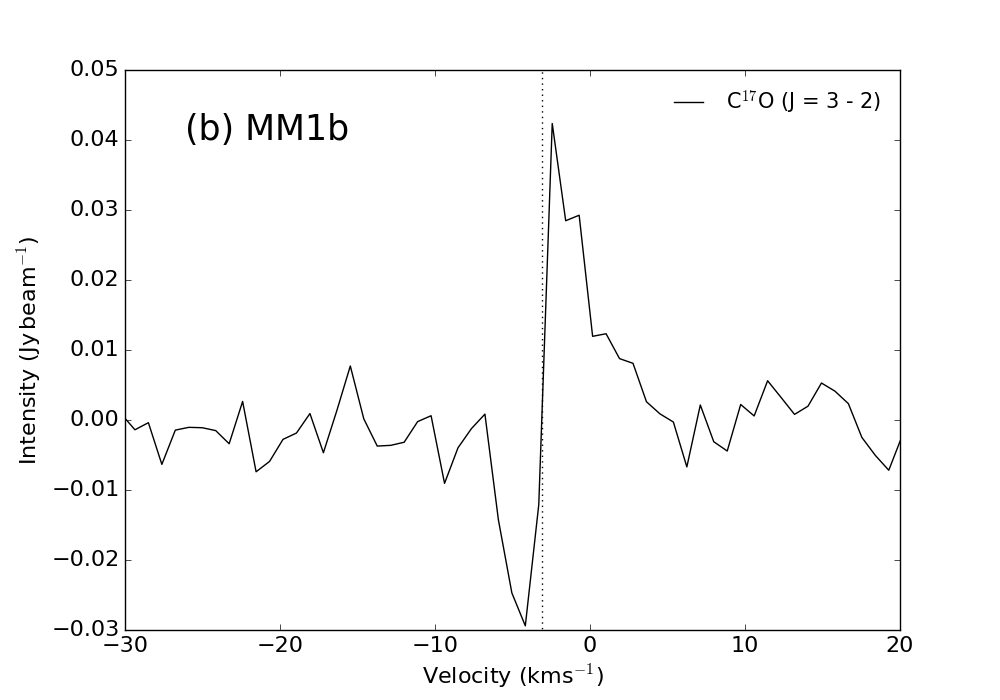}\\
 \includegraphics[width=0.5\textwidth]{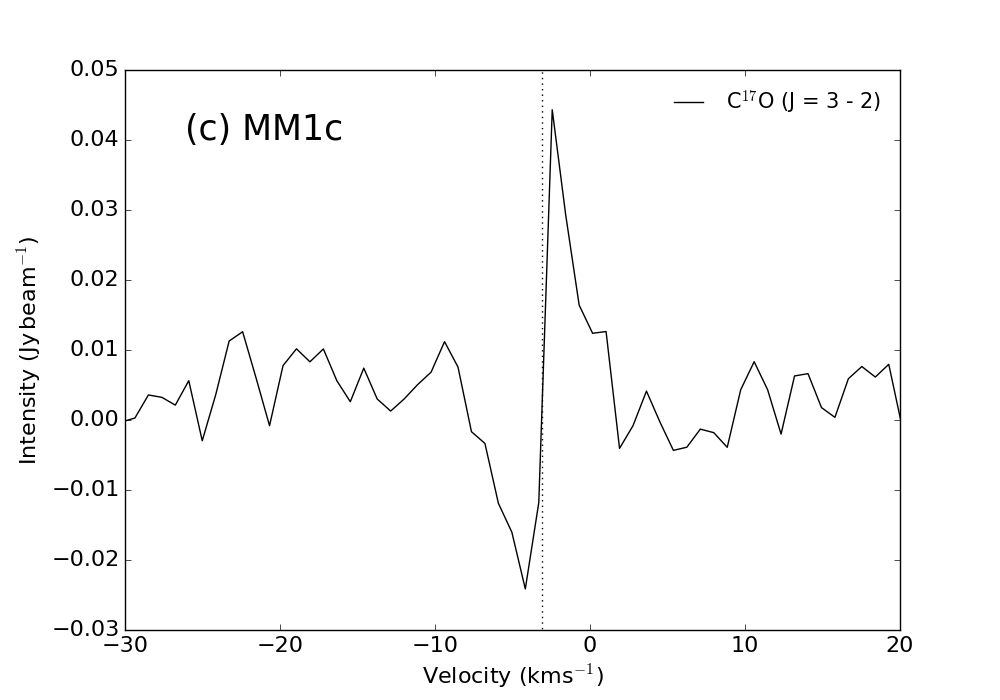} &
 \includegraphics[width=0.5\textwidth]{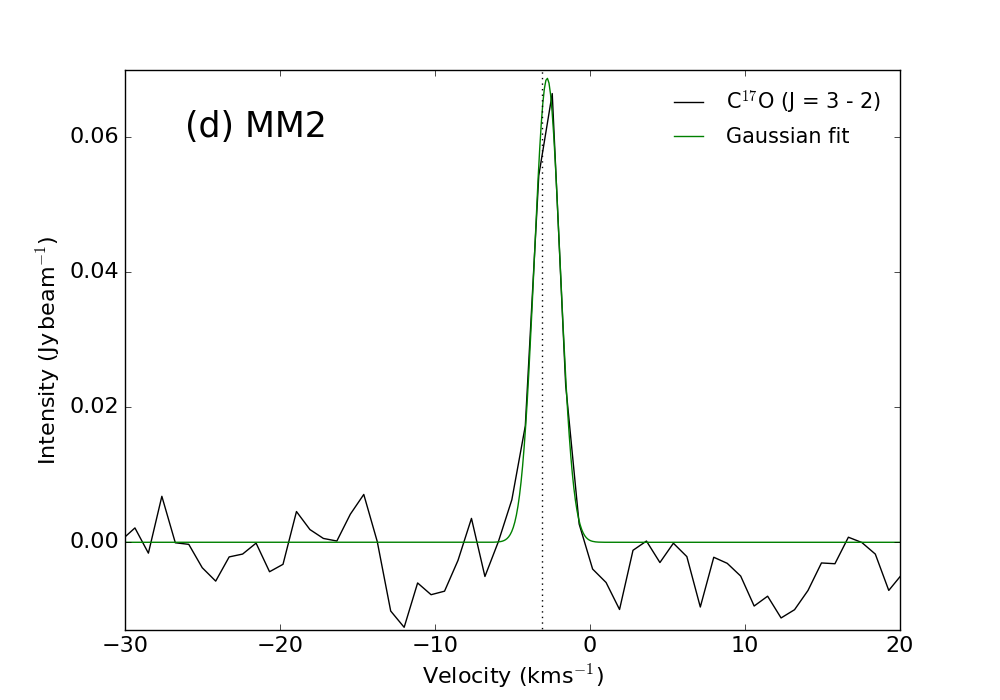}\\
    \end{tabular}
    \caption{C$^{17}$O spectra of: (a) MM1a (b) MM1b (c) MM1c and (d) MM2. Vertical dotted line indicates the systemic velocity of the source.}
    \label{fig:C17O_line}
\end{figure*}

\begin{figure*}
\centering
\begin{tabular}{cl}
 \includegraphics[width=0.5\textwidth]{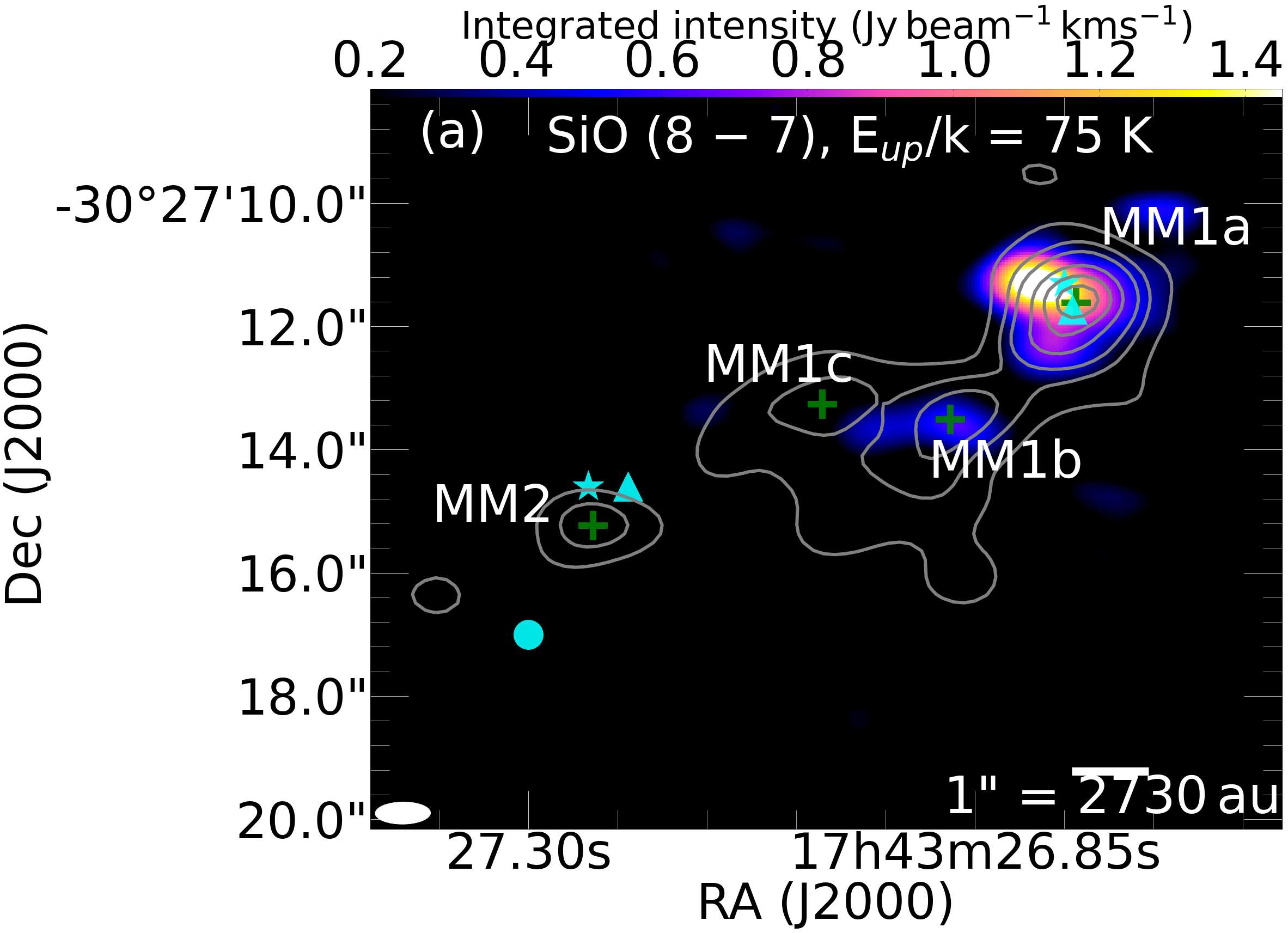} &
 \includegraphics[width=0.5\textwidth]{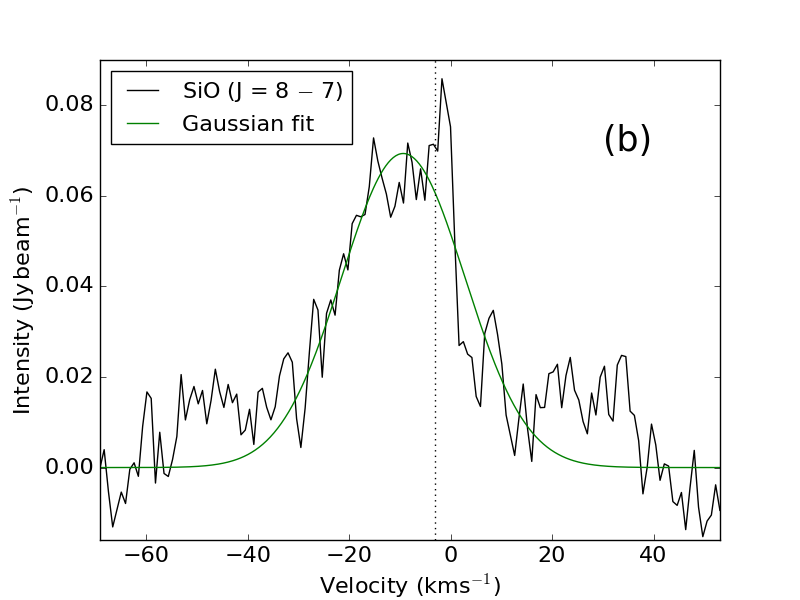}\\
 \end{tabular}
   \caption{SiO line (a) integrated intensity map (background) and (b) spectrum. Gray contours, green crosses, cyan stars, triangles, circle and white ellipse are the same as defined in Fig.\,\ref{fig:c17o_m0}. Vertical dotted line is the same as in Fig.\,\ref{fig:C17O_line}.}
    \label{fig:sio}
\end{figure*}

\begin{figure*}
\centering
	\includegraphics[width=0.95\textwidth]{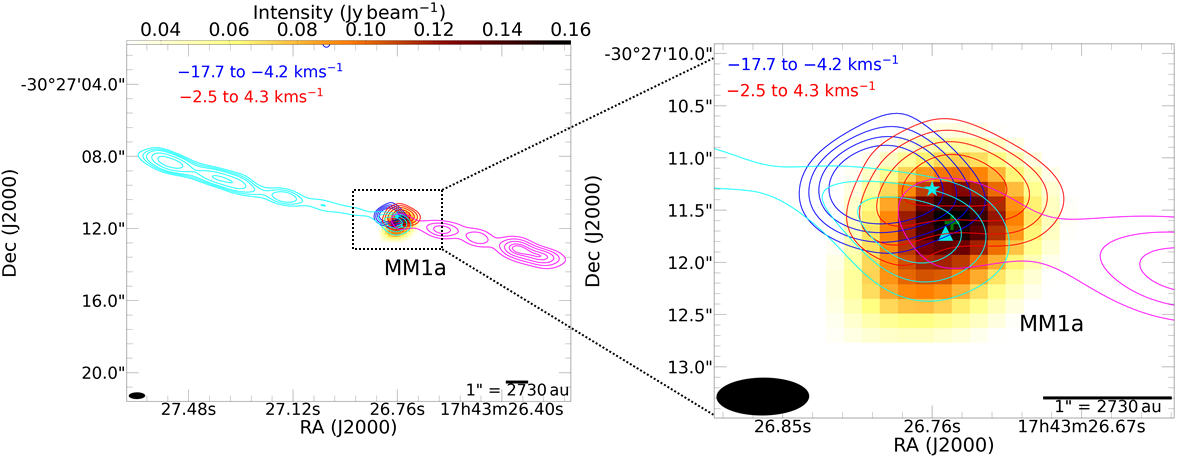} 
    \caption{$Left$: The blue-shifted and red-shifted SiO emission superimposed on the ALMA dust continuum image (background) as blue and red contours, respectively. $Right$: Zoomed-in of the selected region. Cyan and magenta contours indicate the direction of the bipolar outflow traced by $^{12}$CO in MM1a \citep[see][]{2023MNRAS.520.4747U}. Green cross, cyan star and triangle are the same as defined in Fig.\,\ref{fig:c17o_m0}. Black ellipse at the left bottom corner represents the ALMA synthesized beam.}
    \label{fig:sio_outflow}
\end{figure*}

\begin{figure*}
\centering
 \includegraphics[width=0.9\textwidth]{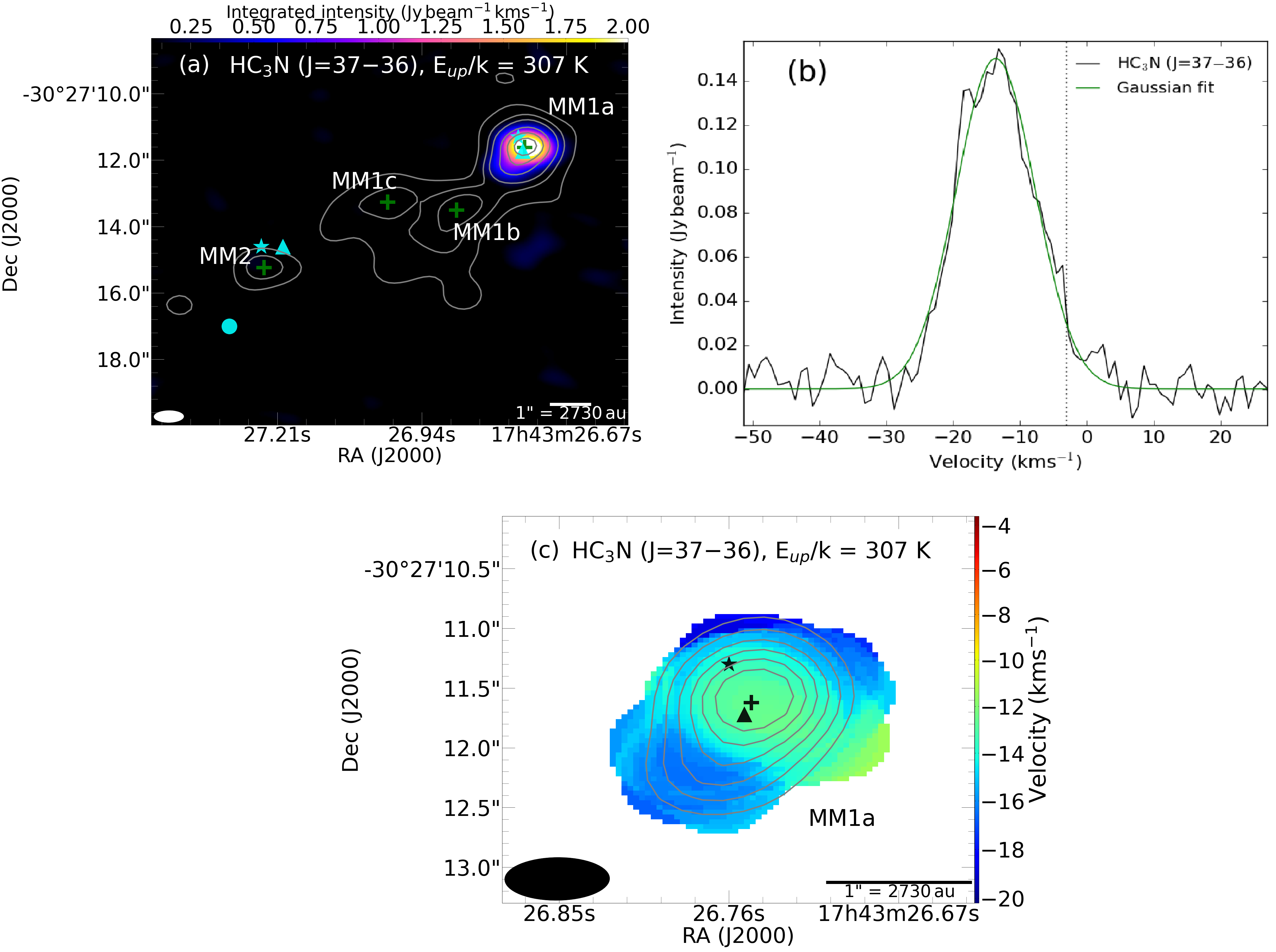}
   \caption{HC$_3$N line (a) integrated intensity map (background), (b) spectrum extracted around the continuum peak of MM1a and (c) velocity field map zoomed in between $-$20.2 to $-$3.6\,kms$^{-1}$ for clarity. Gray contours (levels = 60, 75, 90, 105, 120, 135\,mJy\,beam$^{-1}$) indicate the ALMA dust continuum image. Black cross, star and triangle indicate the peak positions of the dust continuum peak, 6.7\,GHz CH$_3$OH and 22\,GHz H$_2$O masers, respectively. Gray contours, green crosses, cyan stars, triangles, circle and white ellipse are the same as defined in Fig.\,\ref{fig:c17o_m0}. Vertical dotted line is the same as in Fig.\,\ref{fig:C17O_line}.}
    \label{fig:hc3n}
\end{figure*}

\section{Results}
\label{sec:result}
%\lipsum[2]

The spatial distribution of C$^{17}$O, SiO, HC$_{3}$N and SO$_{2}$ molecules compared to the dust continuum cores \citep[MM1a, MM1b, MM1c and MM2; see][]{2023MNRAS.520.4747U} is presented as follow;

\subsection{C$^{17}$O (J = 3 $-$ 2) emission}
\label{sec:c17o}
%\lipsum[3]

The integrated intensity map of C$^{17}$O emission is shown in Fig.\,\ref{fig:c17o_m0}. The C$^{17}$O emission is integrated over a velocity range of $-$10.9 to 6.7\,kms$^{-1}$. The spatial distribution of C$^{17}$O molecule shows an extended emission (2\farcs9, corresponding to 7917\,au at the distance of 2.73 kpc to the source) in MM1a, with an extension (3\farcs4, corresponding to 9282\,au at the distance to the source) to the south, embracing MM1b. The peak emission of the C$^{17}$O molecule coincides with the continuum peak of MM1a. Faint or dim C$^{17}$O emission was found in MM2, with no visible C$^{17}$O emission towards MM1b and MM1c, which could be attributed to low gas column density toward these objects. The C$^{17}$O emission reveals a striking filamentary feature that does not coincide with any of the dust continuum cores and also a dumbbell-shaped feature around a trivial dust object, with barely detectable continuum ($< 3\sigma$), located south-east of MM2. MM1a and MM2 exhibit 4.5\,$\mu$m infrared emission, which is a good indicator of outflows from MYSOs \citep{2008AJ....136.2391C}. MM2 is known to be associated with HII region \citep[see Fig.\,1 in][]{2023MNRAS.520.4747U}. This HII region might be expanding and compressing the gas around it, giving rise to the C$^{17}$O filamentary and dumbbell-shaped features on both sides of MM2.

The C$^{17}$O emission channel map is shown in Fig.\,\ref{fig:C17O_chanmaps}. The channel map was uniformly displaced from $-$7.6 to 1.9\,kms$^{-1}$. The emission started from channel $-$7.6\,kms$^{-1}$, with faint emission and increases upto channel $-$3.3\,kms$^{-1}$, where it peaks and gradually decreases, until it fades out in channel 1.9\,kms$^{-1}$. The blue-shifted emission is dominant from channels $-$7.6 to $-$5.9\,kms$^{-1}$, while faint red-shifted emission appeared from channels $-$1.6 to 1.0\,kms$^{-1}$. Both the blue-shifted and red-shifted emissions overlap from channels $-$5.0 to $-$2.4\,kms$^{-1}$. The emission at channel $-$3.3\,kms$^{-1}$ is consistent with the morphology of the C$^{17}$O shown in Fig.\,\ref{fig:c17o_m0}. The peak in the C$^{17}$O profile of MM1a, close to the systemic velocity \citep[$-$3.1 $\pm$ 0.2\,kms$^{-1}$][]{2023MNRAS.520.4747U} is in agreement with the brightest emission observed in channel $-$3.3\,kms$^{-1}$. The blue-shifted emission located northwest of continuum peak of MM1a and the red-shifted emission found southeast of MM1a are in support of C$^{17}$O tracing a rotating envelope of gas around MM1a.

The C$^{17}$O spectra extracted around the continuum peak of the various cores (MM1a, MM1b, MM1c and MM2) are shown in Fig.\,\ref{fig:C17O_line}. The C$^{17}$O profiles of MM1a (Fig.\,\ref{fig:C17O_line}\,(a)) and MM2 (Fig.\,\ref{fig:C17O_line}\,(d)) show narrow emission with unblended lines. The 1-dimensional Gaussian fit on the C$^{17}$O profiles of MM1a and MM2 show that the emission peaks very close to the systemic velocity of the source, with MM1a (FWHM $\sim$ 5.3 $\pm$ 0.4\,kms$^{-1}$; see Table\,\ref{tab:gaussian}) showing broader line than MM2 (FWHM $\sim$ 3.6 $\pm$ 0.1\,kms$^{-1}$). The C$^{17}$O profile of MM1a reveals presence of double (i.e. blue-shifted and red-shifted) wings, which are associated with outflows and rotating structures. The C$^{17}$O profiles of MM1b (Fig.\,\ref{fig:C17O_line}\,(b)) and MM1c (Fig.\,\ref{fig:C17O_line}\,(c)) reveal the presence of blue-shifted absorption (P Cygni profile within the velocity range of $-$6.8 to $-$3.3\,kms$^{-1}$), which is a sign of expansion or outflow.

\subsection{SiO (J = 8 $-$ 7) emission}
\label{sec:sio}
%\lipsum[4]

The zeroth moment map of SiO molecule calculated from $-$50.5 to 40.2\,kms$^{-1}$ is shown in Fig.\,\ref{fig:sio}\,(a). The SiO emission is highly dominant at 0$\farcs$5 north of MM1a continuum peak position and its peak appears to be offset from the continuum peak of MM1a, suggesting that it is likely associated with activities outside the central core. It is clearly seen that MM1b is associated with weak SiO emission, with no visible emission in MM1c and non-detection (no emission above 3$\sigma$, $\sigma$ = 4.7\,mJy\,beam$^{-1}$) in MM2. The lack of SiO emission in MM1c and weak emission in MM1b could be attributed to the objects (MM1b and MM1c) being in their earliest evolutionary stage and not actively undergoing star formation. Although, MM1b with faint emission is likely to be more actively forming than MM1c. MM2 is probably the most evolved of all the cores, based on its association with HII region and no detectable SiO emission. This is because SiO emission is commonly seen in very young forming cores \citep{2011ApJ...729..124C,2016MNRAS.458.1742C,2020ApJ...896..127K}. The impact of HII region might have led to the dissociation of the SiO molecule in MM2. 

The SiO spectrum extracted within the SiO emission peak, away from the peak position of the central core (MM1a) is shown in Fig.\,\ref{fig:sio}\,(b). The SiO profile reveals broad emission (FWHM $\sim$ 19.4 $\pm$ 0.2\,kms$^{-1}$) and peaks away (with offset velocity of $-$6.9\,kms$^{-1}$) from the systemic velocity of MM1a. The peaking of the SiO emission away from the systemic velocity points to the possibility of SiO tracing event outside the central driving source (MM1a). SiO molecule is an indicator of shock-excited gas from interaction of outflowing gas and the ambient medium. The entirety of the observed SiO emission away from MM1a is probably tracing an outflow-driven shocks from an unresolved YSO in MM1a, which could be indicative of multiplicity of YSOs in MM1a. 

We explored the possibility of SiO emission tracing another bipolar outflow in the source, by extracting SiO emission within the velocity ranges of $-$17.7 to $-$4.2\,kms$^{-1}$ and $-$2.5 to 4.3\,kms$^{-1}$ to form integrated intensities blue and red contours, respectively, which are superimposed on the ALMA dust continuum image (background) as shown in Fig.\,\ref{fig:sio_outflow}. The chosen velocity ranges within the SiO peak emission provide a clear view of the SiO outflow morphology. The blue and red contours indicate the blue-shifted and red-shifted emissions, respectively. The SiO emission shows evidence of another bipolar outflow in the source, with a different outflow morphology compared to the morphology of the $^{12}$CO collimated bipolar outflow previously reported in the source by \citet{2023MNRAS.520.4747U}. It is clearly seen that the blue-shifted and red-shifted SiO emissions are distributed east and west of the position of the 6.7\,GHz CH$_3$OH maser (indicated by cyan star), with the peaks of the blue and red outflow lobes overlapping each other around the peak position of the 6.7\,GHz CH$_3$OH maser. The 6.7\,GHz CH$_3$OH maser is probably connected with the YSO driving the outflow, which is most likely an unresolved YSO in MM1a, requiring higher resolution observations for confirmation. A plausible explanation for the different outflow morphologies is presented in Section\,\ref{sec:orientation}. Also, Fig.\,\ref{fig:sio_outflow} shows that the SiO is more compact than the $^{12}$CO, but extended in the same direction. So the SiO could just be tracing the densest part of the same outflow traced by $^{12}$CO.

\subsection{HC$_{3}$N (J = 37 $-$ 36) emission}
\label{sec:hc3n}
%\lipsum[5]

The zeroth moment map of HC$_{3}$N is shown in Fig.\,\ref{fig:hc3n}\,(a). The HC$_{3}$N emission is integrated over a velocity range of $-$29.2 to 6.5\,kms$^{-1}$. The spatial distribution of HC$_{3}$N molecule reveals an extended emission, with an angular size of $\sim$ 1\farcs6 (which corresponds to $\sim$ 4368\,au at the distance of 2.73\,kpc to the source). The HC$_{3}$N emission corresponds well to the dust continuum distribution of MM1a and peaks at the position of MM1a. Clearly, there is no detectable (no emission above $3\sigma$) HC$_{3}$N emission in MM1b, MM1c and MM2. The lack of HC$_{3}$N emission in MM1b and MM1c could be attributed to the objects (MM1b and MM1c) being in their earliest formative stage and not hot enough to excite the molecule. The reason being that this transition of HC$_{3}$N molecule is a high-excitation line, with a high upper state excitation energy (307\,K) and can only be excited by very hot cores. The lack of HC$_{3}$N emission in these cores (MM1b and MM1c) could also be attributed to low gas density and column density towards these objects. MM2 is associated with HII region and the impact of this region could have led to the dissociation of the HC$_{3}$N molecule in MM2. It is worthy to note that since HC$_{3}$N and SiO have comparable critical densities but very different upper level energies, their relative line strengths probe temperature rather than density differences and comparison with the low critical density tracer C$^{17}$O further constrains column density and optical depth effects.

The HC$_{3}$N spectrum extracted around MM1a continuum peak is shown in Fig.\,\ref{fig:hc3n}\,(b). The HC$_{3}$N profile also reveals broad emission (FWHM $\sim$ 15.5 $\pm$ 0.6\,kms$^{-1}$) with unblended lines and the emission peaking away (with offset velocity of $-$10.5\,kms$^{-1}$) from the systemic velocity of MM1a. The HC$_{3}$N line (FWHM $\sim$ 15.5\,kms$^{-1}$) is broader than the C$^{17}$O line (FWHM $\sim$ 5.3\,kms$^{-1}$), but shows lesser broad line compared to the SiO line (FWHM $\sim$ 19.4\,kms$^{-1}$). The broadening in the HC$_{3}$N profile could be as a result of turbulence near the centre of the YSO (MM1a) driving infall, outflow and rotation (of disk) and as such, the chaos near the centre drives the increase in linewidth (15.5 $\pm$ 0.6\,kms$^{-1}$). The first moment map of HC$_{3}$N (Fig.\,\ref{fig:hc3n}\,(c)) shows a velocity gradient, which is a signature of rotating disk/envelope. It is noteworthy that velocity field map is dominantly blue-shifted, which might be due to global motion towards us or projection effect in which we see mostly the near side.

\subsection{SO$_{2}$ (19$_{1, 19}$ $-$ 18$_{0, 18}$) emission}
\label{sec:so2}
%\lipsum[6]

\begin{figure*}
\centering
 \begin{tabular}{cl}
	\includegraphics[width=0.5\textwidth]{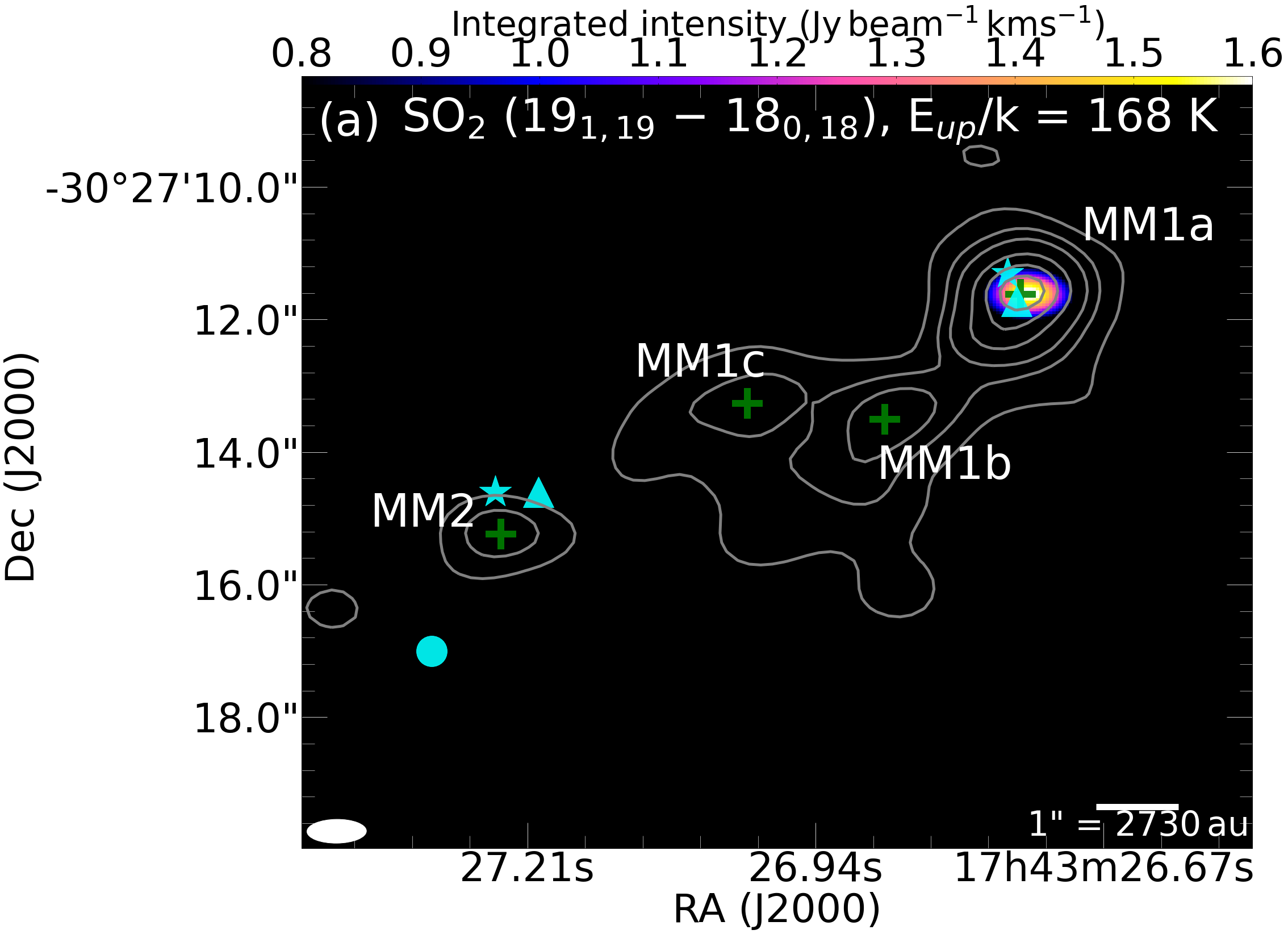} &
 \includegraphics[width=0.5\textwidth]{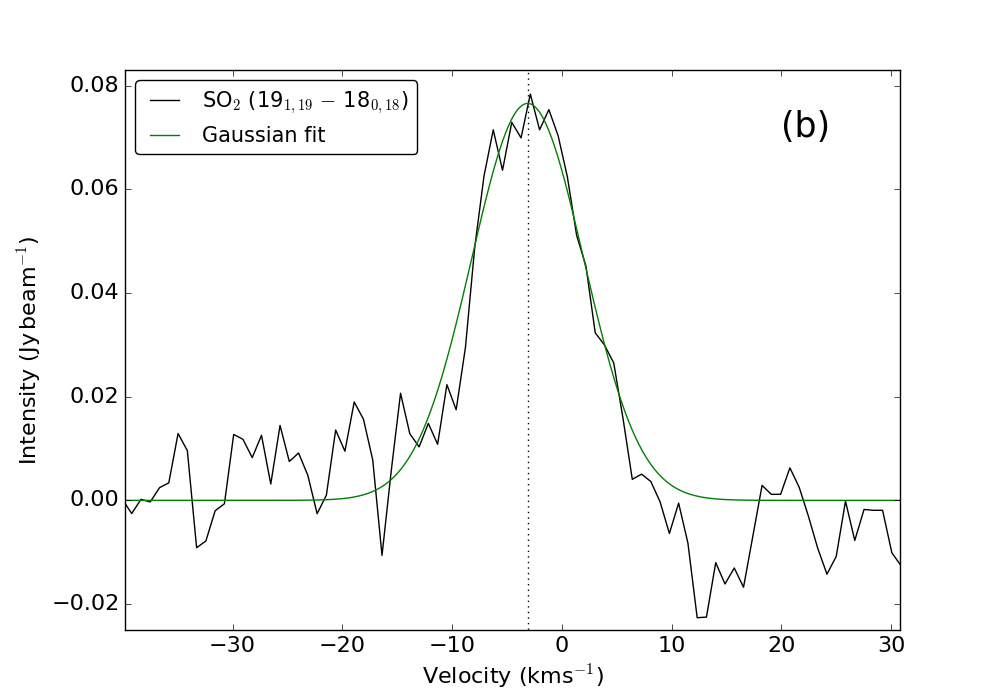}\\ \hspace*{0.000001pt}
 \\
 \includegraphics[width=0.46\textwidth]{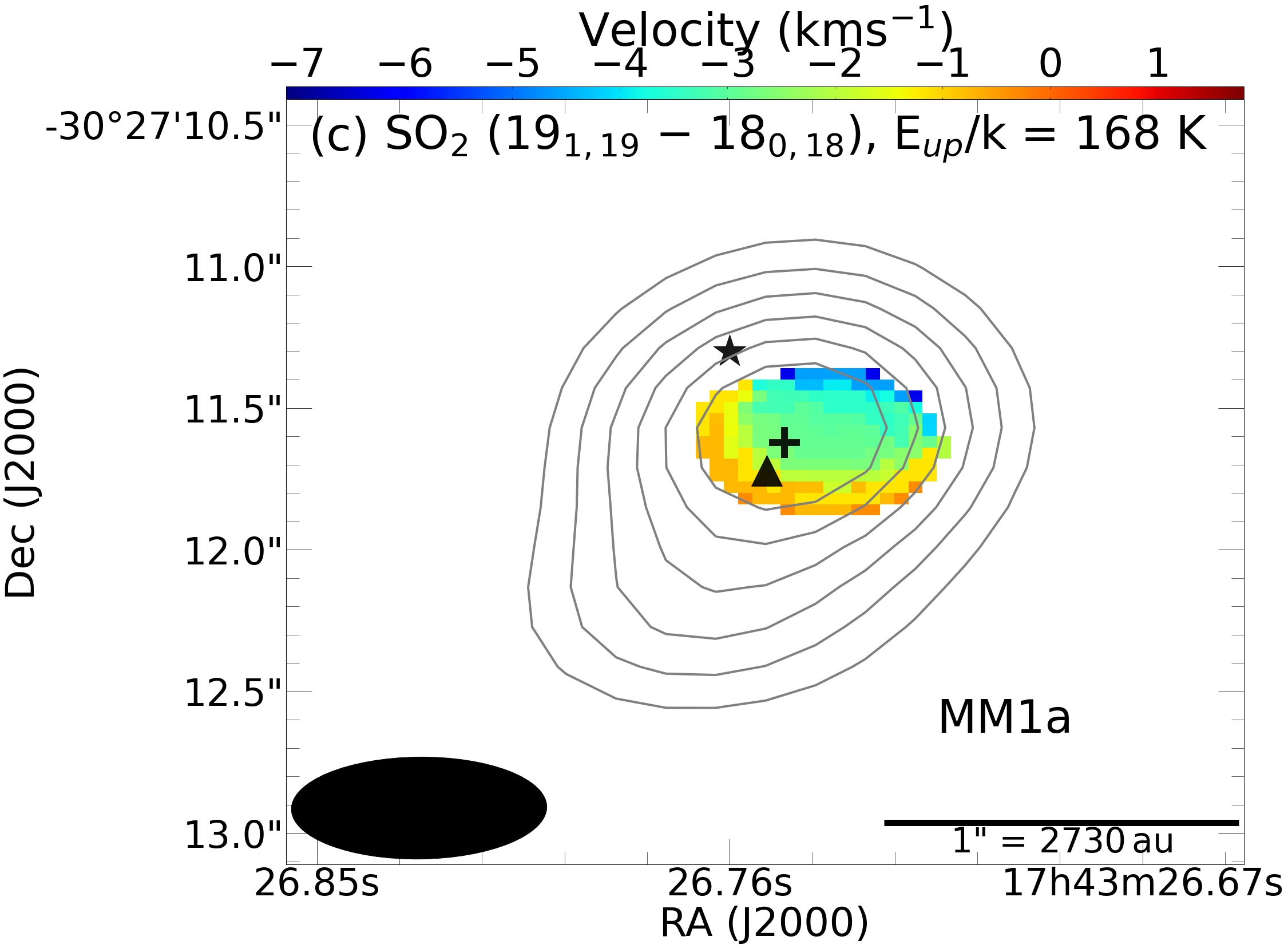} &
 \includegraphics[width=0.46\textwidth]{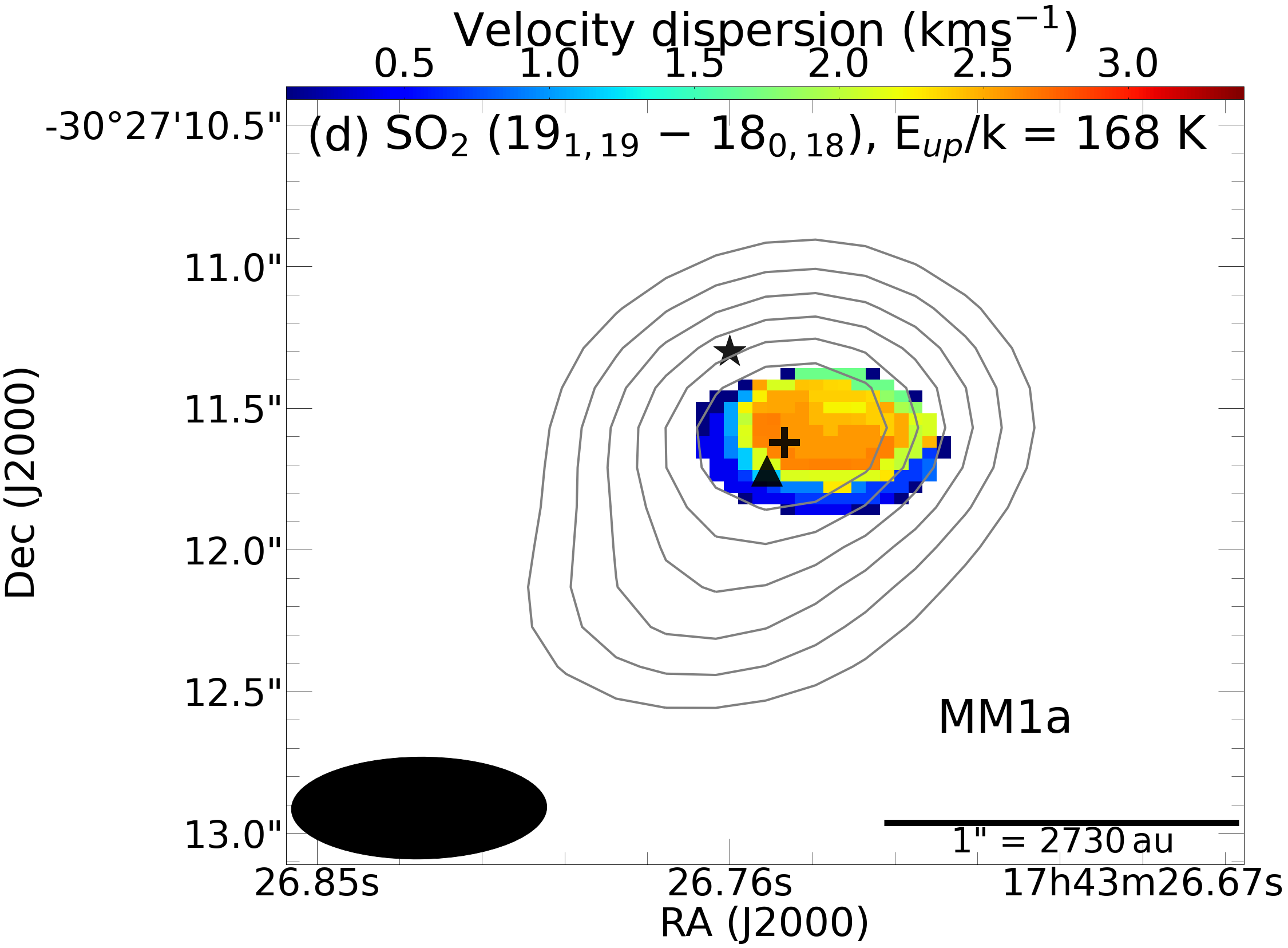}\\ \hspace*{0.000001pt}
    \end{tabular}
    \caption{SO$_{2}$ line (a) integrated intensity map (background), (b) spectrum extracted around the continuum peak of MM1a, (c) velocity field map zoomed in between $-$7.0 to 1.7\,kms$^{-1}$ for clarity and (d) Zoomed in velocity dispersion map. Gray contours, green crosses, cyan stars, triangles, circle and white ellipse are the same as defined in Fig.\,\ref{fig:c17o_m0}. Vertical dotted line is the same as in Fig.\,\ref{fig:C17O_line}. Gray contours, black cross, star, triangle and ellipse are the same as in Fig.\,\ref{fig:hc3n}.}
    \label{fig:so2}
\end{figure*}

The SO$_{2}$ integrated intensity map calculated from $-$17.9 to 11.7\,kms$^{-1}$ is shown in Fig.\,\ref{fig:so2}\,(a). The spatial distribution of SO$_{2}$ molecule reveals a compact morphology in MM1a, with an angular size of $\sim$ 0$\farcs$7 (corresponding to $\sim$ 1911\,au at the distance to the source). The SO$_{2}$ emission appears to reside within the inner core or disk and shows less extended emission, compared to the other emissions (C$^{17}$O, SiO and HC$_{3}$N). The peak of the SO$_{2}$ emission corresponds well with the continuum peak of MM1a. Like the HC$_{3}$N emission, the SO$_{2}$ emission, with a high upper state excitation energy (168\,K) is also not detected (no emission above $3\sigma$) in MM1b, MM1c and MM2. As noted earlier, MM1b and MM1c are not hot enough to excite SO$_{2}$ molecule and the impact of HII region might have led to the dissociation of SO$_{2}$ molecule in MM2, resulting to non-detection. Also, the absence of SO$_{2}$ emission in these cores (MM1b and MM1c) could be as a result of low gas density and column density towards these objects. The SO$_{2}$ spectrum (Fig.\,\ref{fig:so2}\,(b)) extracted around the continuum peak of MM1a shows a profile that peaks at the systemic velocity of the source. The SO$_{2}$ line (FWHM $\sim$ 11.2 $\pm$ 0.4\,kms$^{-1}$) is broader than the C$^{17}$O line (FWHM $\sim$ 5.3 $\pm$ 0.4\,kms$^{-1}$).

The compact morphology of SO$_{2}$ emission suggests a structure experiencing rotational motion centered around the dust continuum peak of MM1a. The SO$_{2}$ emission reveals a velocity gradient (Fig.\,\ref{fig:so2}\,(c)) that is commonly known to be associated with a rotating structure \citep{2016MNRAS.462.4386I,2019AandA...628A...2B,2020ApJ...896...35J,2024Natur.625...55M,2025MNRAS.539.3808U}. The blue-shifted and red-shifted emissions, respectively, extend northwest and southeast of MM1a continuum peak position (black cross), affirming the existence of a rotating structure in MM1a. The zoomed-in velocity dispersion map of SO$_{2}$ emission is shown in Fig.\,\ref{fig:so2}\,(d). The large velocity dispersion ($\sim$ 3\,kms$^{-1}$) within the continuum peak of MM1a is consistent with that of a rotating disk or envelope \citep{2015ApJ...813L..19J,2023MNRAS.525.6146W}. This large velocity dispersion within the inner core of MM1a is a clear indication of active star formation activities (such as outflows and disks or envelopes) in the source.

\section{Discussion}
\label{sec:discussion}
%\lipsum[7]

\subsection{Estimation of outflow properties}
\label{sec:properties}
%\lipsum[8]

\citet{2023MNRAS.520.4747U} previously reported evidence of collimated bipolar outflow traced by $^{12}$CO emission in MM1a. Here, we explore the energetics of MM1a $^{12}$CO and SiO outflows. We calculate the outflow parameters (mass, momentum, energy, dynamical timescale, mass accretion rate, momentum flux and energy ejection rate) using the method described in \citet{2019PASJ...71S...8T}. The mass of outflow, $M_\text{outflow}$ is derived from:
\begin{equation}
M_\text{outflow} = \sum\limits_{|v_{\text{lsr}}-v_{\text{sys}}|\geq2\sigma_{v}} m(v),    
\end{equation}
where $m(v)$ is the mass of each velocity component written as:
\begin{equation*}
m(v) = 4.33 \times 10^{13}\frac{\Bar{\mu}m_{\text{H}}}{\chi}\left[\frac{s(v)}{\textrm{cm$^2$}}\right]f_\tau\left(\frac{T_{\text{ex}}}{\textrm{K}}\right)\text{exp}\left(\frac{5.53}{T_{\text{ex}}}\right)
\end{equation*}
\begin{equation}
\hspace{1.0cm}\times \left[\frac{\overline{T}_{\text{B}}(v)}{\textrm{K}}\right]\left(\frac{\Delta v}{\textrm{km\,s$^{-1}$}}\right), \label{eqn:m}
\end{equation}
In equation (\ref{eqn:m}), $\Bar{\mu}$ is the average molecular weight of 2.4, $m_{\text{H}}$ is the hydrogen atomic mass, $\chi$ is the molecular abundance relative to H$_2$ adopted as 10$^{-4}$ for $^{12}$CO \citep{2009ApJ...696...66Q} and 10$^{-7}$ for SiO \citep{1997AandA...321..293S}, $s(v)$ is the projected area above $3\sigma$ level, $f_\tau = \tau/(1 - e^{-\tau})$ is the correction factor for the optical depth, $\tau = -\text{ln}(1 - \overline{T}_{\text{B}}(v)/T_\text{ex})$, $\overline{T}_{\text{B}}(v)$ is the averaged brightness temperature of all pixels in $s(v)$, $T_{\text{ex}}$ is the excitation temperature of the line and $\Delta v$ is the linewidth. The peak intensities (in Kelvin) of the $^{12}$CO and SiO profiles were obtained from 2-dimensional Gaussian fitting and used as the excitation temperatures of $^{12}$CO (27\,K) and SiO (2.3\,K). The projected area, $s(v)$ is expressed as:
\begin{equation}
s(v) = n_{\text{pix}}\left[1.5 \times 10^{13}\left(\frac{D}{\text{pc}}\right)\left(\frac{\Delta\theta}{\text{arcsec}}\right)\right]^2 \,[\text{cm}^{2}],    
\end{equation}
where $n_{\text{pix}}$ is the pixels number in $s(v)$, $D$ is the source distance and $\Delta\theta$ is the size of pixel = $0\farcs04$. We calculated and adopted a recent near kinematic distance of 2.73 $\pm$ 0.28\,kpc to the source based on the updated kinematic distance estimation method of \citet{2022AJ....164..133R}. The outflow momentum and energy, $P_{\text{outflow}}$ and $E_{\text{outflow}}$ are derived from:
\begin{equation}
P_\text{outflow} = \sum\limits_{|v_{\text{lsr}}-v_{\text{sys}}|\geq2\sigma_{v}} m(v)|v_{\text{lsr}}-v_{\text{sys}}| 
\end{equation}
and
\begin{equation}
E_\text{outflow} = \frac{1}{2}\sum\limits_{|v_{\text{lsr}}-v_{\text{sys}}|\geq2\sigma_{v}} m(v)|v_{\text{lsr}}-v_{\text{sys}}|^2.    
\end{equation}
We derived the inclination angles, $\theta$ of the $^{12}$CO and SiO outflows using the method of \citet{2011MNRAS.413...71K}. This method depends mainly on the accurate location of the driving source. In our calculation, we chose the continuum peak position of MM1a as the driving source and derived inclination angles of $\sim$ 47 $\pm$ 2$^\circ$ and 7 $\pm$ 1 $^\circ$ for the $^{12}$CO and SiO outflows, respectively. These inclination angles are determined with respect to the plane of the sky from this relation:
\begin{equation}
A(\theta) = 1,    
\end{equation}
derived from 
\begin{equation}
A(\theta) = \frac{1 - P}{Q}\frac{1}{\cos\theta} - P\tan\theta\tan\theta,        
\end{equation}
where $P$ and $Q$ are the angle constants, which are, respectively, defined as
\begin{equation}
P = \frac{\beta_1}{\beta_2}    
\end{equation}
and
\begin{equation}
Q = \tan\alpha,    
\end{equation}
where $\beta_1$ and $\beta_1$ are the angle distances in the outflow lobes and $\alpha$ is the angle between the driving source and the axis points on the plane of the sky. The outflow momentum varies with $1/\sin\theta$ and as such, $P\sin\theta$ was effectively calculated. 

The outflow dynamical timescale, $t_\text{d}$ is obtained from:
\begin{equation}
t_\text{d} = \frac{R_{\text{max}}}{\Delta v_{\text{max}}},    
\end{equation}
where $R_{\text{max}}$ is the maximum size of outflow from integrated intensity map and $\Delta v_{\text{max}}$ is the maximum velocity of outflow obtained from $|v_{\text{max}}-v_{\text{sys}}|$, where $v_{\text{max}}$ and $v_{\text{sys}}$ are the highest velocity for emission above $3\sigma$ and systemic velocity of the source, respectively. The mass outflow rate, $\dot{M}_{\text{outflow}}$, momentum outflow rate, $\dot{P}_{\text{outflow}}$ and energy outflow rate, $\dot{E}_{\text{outflow}}$ are calculated from:
\begin{equation}
\dot{M}_{\text{outflow}} = \frac{M_{\text{outflow}}}{t_\text{d}},    
\end{equation}
\begin{equation}
\dot{P}_{\text{outflow}} = \frac{P_{\text{outflow}}}{t_\text{d}}    
\end{equation}
and
\begin{equation}
\dot{E}_{\text{outflow}} = \frac{E_{\text{outflow}}}{t_\text{d}}.    
\end{equation}
The results of the derived outflow parameters for $^{12}$CO and SiO outflows are presented in Table\,\ref{tab:outflow}. The calculated $^{12}$CO outflow values, especially the momentum outflow rate ($\dot{P}_{\text{outflow}}$), is an order of magnitude higher than the largest values obtained in the Orion Molecular Cloud (OMC), whereas the values estimated for the SiO outflow are two orders of magnitude higher than the largest values reported in OMC \citep{2008ApJ...688..344T}, but agree with values reported for powerful SiO outflow sources \citep{2023AandA...677A.148G}. Our estimated values for $M_\text{outflow}$, $P_\text{outflow}$, $E_\text{outflow}$, $\dot{M}_{\text{outflow}}$, $\dot{P}_{\text{outflow}}$ and $\dot{E}_{\text{outflow}}$ are in agreement with simulations \citep{2017MNRAS.470.1026M,2020AJ....160...78R} and large scale surveys \citep{2014MNRAS.444..566D,2015MNRAS.453..645M} of molecular outflows in massive star forming regions (MSFRs). The dynamical timescale, $t_{\text{d}}$ we derived ($\sim$ 10$^{3}$\,yr) is much lower than $t_{\text{d}}$ for most molecular outflows observed in the above-mentioned surveys, where $t_{\text{d}} \sim 10^4 - 10^5$\,yr, but agrees with the smallest values obtained in OMC and G11.92$-$0.61/G19.01$-$0.03 \citep{2011ApJ...729..124C}. The derived radio luminosity of the driving source is $\sim$ 8.5 $\times$ 10$^{-3}$$L_{\odot}$, which is far less than the bolometric luminosity ($\sim$ 1.15 $\times$ 10$^{4}$$L_{\odot}$) as expected for a massive protostar. Our mass and momentum outflow rates are consistent with the $\dot{M}_{\text{outflow}}$ and $\dot{P}_{\text{outflow}}$ expected for a driving source of luminosity $\sim 10^4L_{\odot}$ \citep[$\dot{M}_{\text{outflow}}$ $\sim$ 10$^{-5}$ to 10$^{-3}$\,$M_{\odot}$\,yr$^{-1}$ and $\dot{P}_{\text{outflow}}$ $\sim$ 10$^{-3}$ to 10$^{-1}$\,$M_{\odot}$\,\kms\,yr$^{-1}$;][]{1992AandA...261..274C,2002AandA...383..892B,2005ApJ...625..864Z}. These powerful outflows suggest a higher mass accretion rate ($\sim$ 10$^{-3}$\,$M_{\odot}$\,yr$^{-1}$) than what might be expected for a single protostar, which could imply the presence of multiple driving sources. Our results indicate that the driving sources of $^{12}$CO and SiO outflows in G358.46$-$0.39 are at least two separate massive protostars.

\begin{table}
\centering
\caption{Properties of the observed lines}
\begin{tabular}{lccr}
\hline
Molecular lines$^*$ & Frequency & $E_{\text{up}}/k$ & $\rho_{\text{c}}$ \\
{} & (GHz) & (K) & (cm$^{-3}$) \\
\hline
C$^{17}$O (J = 3 $-$ 2) & 337.061 & 32 & 6.6 $\times$ 10$^{4}$ \\
SiO (J = 8 $-$ 7) & 347.330 & 75 & 4.8 $\times$ 10$^{7}$ \\
HC$_{3}$N (J = 37 $-$ 36) & 336.520 & 307 & 6.1 $\times$ 10$^{7}$ \\
SO$_{2}$ (19$_{1,19}$ $-$ 18$_{0,18}$) & 346.652 & 168 & 1.0 $\times$ 10$^{7}$ \\
\hline
\end{tabular}
\vspace{1.5mm}
\begin{minipage}{8cm}
Notes.\\
$^*$Line parameters are from the Cologne Database for Molecular Spectroscopy\footnote{{\rm https://cdms.astro.uni-koeln.de/cgi-bin/cdmssearch}} \citep[CDMS;][]{2001AandA...370L..49M,2005JMoSt.742..215M,2016JMoSp.327...95E}.\\ 
$E_{\text{up}}/k$ is the upper state energy.\\ 
$\rho_{\text{c}}$ is the critical density.
\end{minipage}
\label{tab:properties}
\end{table}

%\begin{table}
%\centering
 %   \caption{Properties of the observed lines}
%	\begin{tabular}{lccr}
 %       \hline
  %      Molecular lines &  Frequency & $E_{\text{up}}/k$ & $\rho_{\text{c}}$ \\
   %     {}  & (GHz) & (K) & (cm$^{-3}$)\\
    %    \hline
     %  C$^{17}$O (J = 3 $-$ 2) & 337.061 & 32 & 6.6 $\times$ 10$^{4}$\\
      % SiO (J = 8 $-$ 7) & 347.330 & 75 & 4.8 $\times$ 10$^{7}$\\
       %HC$_{3}$N (J = 37 $-$ 36) & 336.520 & 307 & 6.1 $\times$ 10$^{7}$\\
       %SO$_{2}$ (19$_{1, 19}$ $-$ 18$_{0, 18}$) & 346.652 & 168 & 1.0 $\times$ 10$^{7}$ \\
    %\hline
    %\multicolumn{4}{@{Notes.}l@{}}% 
%{}\\
%\multicolumn{4}{@{}l@{}}% 
%{$E_{\text{up}}/k$ is the upper state energy.}\\
%\multicolumn{4}{@{}l@{}}% 
%{$\rho_{\text{c}}$ is the critical density.}\\
%\multicolumn{4}{@{}l@{}}% 
%\multicolumn{4}{l}{%
%  \begin{minipage}{6.5cm}%
%$^b$ Cologne Database for Molecular Spectroscopy\footnote{https://cdms.astro.uni-koeln.de/cgi-bin/cdmssearch} \citep[CDMS;][]{2001AandA...370L..49M,2005JMoSt.742..215M,2016JMoSp.327...95E}}.
%  \end{minipage}%
%}
%\begin{minipage}{10.5cm}
%{Cologne Database for Molecular Spectroscopy\footnote{https://cdms.astro.uni-koeln.de/cgi-bin/cdmssearch} \citep[CDMS;][]{2001AandA...370L..49M,2005JMoSt.742..215M,2016JMoSp.327...95E}.}\\
%\end{minipage}
%    \label{tab:properties}
 %   \end{tabular}
%\end{table}

\begin{table}
\centering
    \caption{Results of the Gaussian fits to the line profiles of MM1a}
	\begin{tabular}{lccr}
        \hline
        Molecules & Intensity & V$_{\text{LSR}}$ & FWHM \\
        {}  & (Jy\,beam$^{-1}$) & (kms$^{-1}$) & (kms$^{-1}$)\\
        \hline
       C$^{17}$O (J = 3 $-$ 2) & 0.181$\pm$0.005 & $-$3.3$\pm$0.1 & 5.3$\pm$0.4\\
       SiO (J = 8 $-$ 7) & 0.068$\pm$0.002 & $-$10.0$\pm$0.7 & 19.4$\pm$0.2\\
       HC$_{3}$N (J = 37 $-$ 36) & 0.149$\pm$0.008 & $-$13.6$\pm$0.1 & 15.5$\pm$0.6\\
       SO$_{2}$ (19$_{1, 19}$ $-$ 18$_{0, 18}$) & 0.078$\pm$0.003 & $-$3.1$\pm$0.2 & 11.2$\pm$0.4\\
    \hline
\multicolumn{4}{@{Notes.}l@{}}% 
{}\\
\multicolumn{4}{@{}l@{}}% 
{V$_{\text{LSR}}$ is the velocity local standard of rest.}\\
\multicolumn{4}{@{}l@{}}% 
{FWHM is the full width at half maximum.}\\
    \label{tab:gaussian}
    \end{tabular}
\end{table}

\begin{table*}
\centering
    \rotatebox{90}{
    \begin{minipage}{\textheight}
    \caption{Derived outflow parameters}
	\begin{tabular}{lccccccccr}
        \hline
        Outflow lobe & $|v_{\text{max}}-v_{\text{sys}}|$ & $M_\text{outflow}$ & $P_\text{outflow}$ & $E_\text{outflow}$ &  $R_{\text{max}}$ & $t_\text{d}$ & $\dot{M}_{\text{outflow}}$ & $\dot{P}_{\text{outflow}}$ & $\dot{E}_{\text{outflow}}$\\
        {}  & (kms$^{-1}$) & ($M_{\odot}$) & ($M_{\odot}$\,\kms) & (erg) & (pc) & (yr) & ($M_{\odot}$\,yr$^{-1}$) & ($M_{\odot}$\,\kms\,yr$^{-1}$) & (erg\,s$^{-1}$)\\
        \hline
        {} & {} & {} & $^{12}$CO (J = 3 $-$ 2) & {} & {} & {} & {} & {} & {}\\
        \hline
        Blue & 46.9 & 1.92$\pm$0.02 & 58.6$\pm$5.6 & 3.2$\pm$0.3\,$\times$\,10$^{46}$ & 0.201$\pm$0.021 & 4.2$\pm$0.4\,$\times$\,10$^{3}$ & 4.58$\pm$0.05\,$\times$\,10$^{-4}$ & 1.4$\pm$0.1\,$\times$\,10$^{-2}$ & 2.4$\pm$0.2\,$\times$\,10$^{35}$\\ 
        Red & 
        45.4 & 0.666$\pm$0.007 & 19.7$\pm$2.9 & 1.04$\pm$0.15\,$\times$\,10$^{46}$ & 0.106$\pm$0.011 & 2.3$\pm$0.2\,$\times$\,10$^{3}$ & 2.92$\pm$0.03\,$\times$\,10$^{-4}$ & 0.9$\pm$0.1\,$\times$\,10$^{-2}$ & 1.4$\pm$0.2\,$\times$\,10$^{35}$\\ 
        Average & {} & 1.295$\pm$0.014 & 39.2$\pm$4.2 & 2.11$\pm$0.23\,$\times$\,10$^{46}$ & {} & 3.2$\pm$0.3\,$\times$\,10$^{3}$ & 3.75$\pm$0.04\,$\times$\,10$^{-4}$ & 1.1$\pm$0.1\,$\times$\,10$^{-2}$ & 1.9$\pm$0.2\,$\times$\,10$^{35}$\\ 
        \hline
        {} & {} & {} & SiO (J = 8 $-$ 7) & {} & {} & {} & {} & {} & {}\\
        \hline
         Blue & 12.1 & 172.6$\pm$1.8 & 195.2$\pm$28.8 & 1.6$\pm$0.2\,$\times$\,10$^{47}$ & 0.016$\pm$0.002 & 1.1$\pm$0.1\,$\times$\,10$^{3}$ & 162.2$\pm$1.7\,$\times$\,10$^{-3}$ & 18.3$\pm$2.7\,$\times$\,10$^{-2}$ & 4.7$\pm$0.7\,$\times$\,10$^{36}$\\ 
        Red & 4.8 & 61.1$\pm$0.6 & 32.9$\pm$3.1 & 1.4$\pm$0.1\,$\times$\,10$^{46}$ & 0.019$\pm$0.002 & 2.5$\pm$0.3\,$\times$\,10$^{3}$ & 24.9$\pm$0.3\,$\times$\,10$^{-3}$ & 1.3$\pm$0.1\,$\times$\,10$^{-2}$ & 1.8$\pm$0.2\,$\times$\,10$^{35}$\\
        Average & {} & 116.9$\pm$1.2 & 114.1$\pm$15.9 & 8.7$\pm$1.2\,$\times$\,10$^{46}$ & {} & 1.8$\pm$0.2\,$\times$\,10$^{3}$ & 93.6$\pm$1.0\,$\times$\,10$^{-3}$ & 9.8$\pm$1.4\,$\times$\,10$^{-2}$ & 2.5$\pm$0.4\,$\times$\,10$^{36}$\\
        \hline
        \label{tab:outflow}
	\end{tabular}
    \end{minipage}
}
\end{table*}

\begin{figure*}
\centering
	\includegraphics[width=0.5\textwidth]{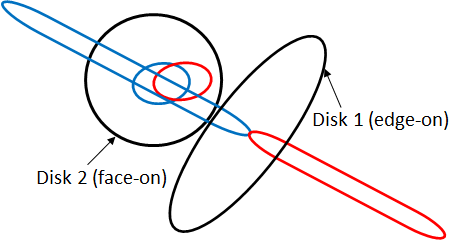}
    \caption{Illustration of disks orientations of probably two separate cores in MM1a.}
    \label{fig:schematic}
\end{figure*}

\subsection{Disk orientation in G358.46$-$0.39}
\label{sec:orientation}
%\lipsum[9]

 G358.46$-$0.39 is an extended green object \cite[EGO;][]{2013ApJS..206...22C} and EGOs are used to study outflows in MSFRs \cite{2008AJ....136.2391C}. EGOs are known to have a central knot (disk) and lobes, which represent the outflowing materials or ejections. Depending on their disk orientation, the outflow lobes can either be spatially overlapped or separated for a disk viewed face-on or egde-on, respectively. The disk orientation in the source was investigated. We used the ratio of major and minor axis lengths of EGO G358.46$-$0.39 from Spitzer 4.5\,$\mu$m image \citep{2003PASP..115..953B} to determine the full width and length of the clump outflow. The major and minor axis lengths of G358.46$-$0.39 from Spitzer 4.5\,$\mu$m image are 8\farcs2 and 7\farcs7, respectively. The ratio of the lengths of the major and minor axes of the clump is found to be 1.1. The compact morphology of EGO G358.46$-$0.39 implies a possible orientation where the outflow disk is observed face-on. Looking at the source morphology and axis ratio, G358.46$-$0.39 is similar or different to the examples of \citet{2024MNRAS.530.1956S}. 

 According to \citet{2021MNRAS.507.1138M}, periodic methanol maser sources with light curves close to the absolute sine function are likely to show face-on disk orientation. G358.46$-$0.39 is a periodic methanol maser source and has a light curve that shows sinusoidal flux variation. As such, it is expected to exhibit a face-on disk orientation. The overlap in the peaks of the blue and red lobes of the SiO outflow is an indication that the outflow direction is along the line of sight of the observer, implying that the apparent disk of the YSO driving the outflow has a face-on geometry. The slight shift between the blue and red lobes of the SiO outflow is due to slight outflow inclination ($\sim$ 7$^\circ$) with respect to the line of sight. However, a clear separation was observed between the blue and red lobes of $^{12}$CO outflow reported in \citet{2023MNRAS.520.4747U}, indicating that the outflow direction is almost perpendicular to the line of sight of the observer, which implies that the apparent disk of the YSO has an edge-on geometry. One plausible explanation for these different outflow morphologies in G358.46$-$0.39 is that MM1a hosts at least two unresolved YSOs driving outflows. The central YSO driving the $^{12}$CO outflow associated with a disk viewed edge-on, while the second YSO driving the SiO outflow has a disk viewed face-on, as illustrated in Fig.\,\ref{fig:schematic}. It will be of great importance to carry out higher resolution observations to confirm this scenario.

\section{Conclusions}
\label{sec:conclusion}
%\lipsum[1-10]

We have investigated the gas distribution of C$^{17}$O, SiO, HC$_{3}$N and SO$_{2}$ molecules, as well as the energetics of $^{12}$CO and SiO outflows in G358.46$-$0.39 proto-cluster employing ALMA band 7 archival data, with the aim of providing an improved understanding of its protostellar nature, gas kinematics and dynamics. The integrated intensity map of C$^{17}$O reveals filamentary and dumbbell-shaped structures that are probably compressed gases from the expansion of the HII region MM2. The SiO emission reveals spatially overlapped blue and red outflow lobes, likely driven by an unresolved YSO in MM1a. The spatial distribution of HC$_3$N and SO$_{2}$ molecules in MM1a reveals a compact morphology, with no visible HC$_3$N and SO$_{2}$ emissions in the other cores (MM1b, MM1c and MM2). The velocity field map of SO$_{2}$ emission shows a clear velocity gradient that is commonly seen in rotating structures. The large velocity dispersion ($\sim$ 3\,\kms) seen within the continuum peak of MM1a is consistent with that of a rotating disk or envelope.

We also estimated the average mass, momentum and energy outflow rate, as well as other outflow parameters, which are consistent with simulations and observations of molecular outflows in MSFRs. The SiO outflow exhibits a different morphology compared to the $^{12}$CO outflow morphology previously observed in MM1a. The SiO and $^{12}$CO outflows are probably associated with disks of separate cores with one face-on and the other edge-on, pointing to at least two driving YSOs in MM1a. The gas motion and energetics of outflows in MM1a indicate that it is a massive protostar that is actively accreting and undergoing star formation. MM1a is considered to be a good site for testing some fundamental theory of star formation and this work has contributed to the growing statistics of well-studied MYSOs.

\section*{Acknowledgements}
CJU and WO acknowledge financial support from the University of South Africa (Research Fund: 409000). JOC is supported by the Italian Ministry of Foreign Affairs and International Cooperation (MAECI Grant Number ZA18GR02) and the South African Department of Science and Technology's National Research Foundation (DST-NRF Grant Number 113121) as part of the ISARP RADIOSKY2020 Joint Research Scheme. This paper makes use of the following ALMA data: ADS/JAO.ALMA\#2013.1.00960.S. ALMA is a partnership of ESO (representing its member states), NSF (USA) and NINS (Japan), together with NRC (Canada) and NSC and ASIAA (Taiwan), in cooperation with the Republic of Chile. The Joint ALMA Observatory is operated by ESO, AUI/NRAO and NAOJ. %\cite{Aladro15}

%\appendix

%\section{Appendix title 1}

%\section{Appendix title 2}

\bibliographystyle{elsarticle-harv} 
\bibliography{g358}


%\begin{thebibliography}{00}

%% For example:

 %\bibitem[Aladro et al.(2015)]{Aladro15} Aladro, R., Martín, S., Riquelme, D., et al. 2015, aas, 579, A101

%\end{thebibliography}

\end{document}